\documentstyle[sprocl,epsf]{article}

\def\etal{{\it et al.\ }}
\def\lsim{\raise0.3ex\hbox{$<$}\kern-0.75em{\lower0.65ex\hbox{$\sim$}}} 
\def\gsim{\raise0.3ex\hbox{$>$}\kern-0.75em{\lower0.65ex\hbox{$\sim$}}} 

\begin{document}

\title{COSMIC SHOCK WAVES ON LARGE SCALES OF THE UNIVERSE}

\author{DONGSU RYU}
\address{Dept. of Astronomy, Univ. of Washington, Seattle, WA 98195-1580\\
Dept. of Astronomy \& Space Science, Chungnam Nat. Univ., Korea}
\author{HYESUNG KANG}
\address{Dept. of Astronomy, Univ. of Washington, Seattle, WA 98195-1580}

\maketitle\abstracts{
In the standard theory of the large scale structure formation, matter
accretes onto high density perturbations via gravitational instability. 
Collisionless dark matter forms caustics around such structures, while 
collisional baryonic matter forms accretion shocks which then halt and
heat the infalling gas.
Here, we discuss the characteristics, roles, and observational
consequences of these accretion shocks.}

The simulations of large scale structure in the universe, which include
the evolution of baryonic matter as well as that of dark matter, have
shown the formation of accretion shocks around the nonlinear structures
such as supergalactic sheets, filaments, and clusters of galaxies (see,
for example, Kang \etal 1994).
The upper panel of Figure shows the density contours of baryonic matter
in one of those simulations (Kulsrud \etal 1997; Ryu, Kang, \& Biermann 1997).
Accretion shocks exist around the high density structures of clusters,
filaments, and sheets in the density contours.

The properties of the shocks and the accreting matter outside the shocks
depend upon the power spectrum of the initial perturbations on a given
scale as well as the background expansion in a given cosmological model.
To study them, we calculated the accretion of dark matter particles
around clusters in one-dimensional spherical geometry under various
cosmological models (Ryu \& Kang 1997).
The velocity of the accreting matter around clusters of a given temperature
is smaller in a universe with smaller $\Omega_o$, but only by up to
$\sim24\%$ in the models with $0.1\le \Omega_o \le 1$.
It is given as $v_{acc}\approx0.9-1.1\times10^3{\rm km~s^{-1}}
[(M_{cl}/R_{cl})/(4\times10^{14}{\rm M}_{\odot}/{\rm Mpc})]^{1/2}$.
However, the accretion velocity around clusters of a given mass or 
a given radius depends more sensitively on the cosmological models.

Considering that these accretion shocks are very big with a typical size
$\gsim$ a few Mpc and very strong with a typical velocity jump $\gsim$
a few $1000~{\rm km}~{\rm s}^{-1}$, they could serve as possible sites
for the acceleration of high energy cosmic rays by the first-order
Fermi process (Kang, Ryu, \& Jones 1996; Kang, Rachen, \& Biermann 1997).
With Jokipii diffusion, the observed cosmic ray spectrum near $10^{19}$eV
could be explained with reasonable parameters if about $10^{-4}$ of the
infalling kinetic energy can be injected into the intergalactic space as
the high energy particles.

The shocks could serve also as sites for the generation of weak seeds
of cosmic magnetic field by the Biermann battery mechanism.
Then, these seeds could be amplified to strong (up to a few $\mu$G)
and coherent (up to the galaxy scale) magnetic field by the Kolmogoroff
turbulence endemic to gravitational structure formation (Kulsrud \etal 1997).
The lower panel of Figure shows the vectors of seed magnetic field
generated by the Biermann mechanism in the simulation.
In the highest density regions of clusters, the magnetic field is chaotic
since the flow motion is turbulent.
However, in the regions which are identified as filaments or
sheets, the magnetic field is aligned with the structures due to the
streaming flow motion along the structures.

If there is aligned magnetic field in filaments and sheets, this would
induce the Faraday rotation in polarized radio waves from extra-galactic
sources.
Then, an upper limit in its strength can be placed by comparing the
expected rotational measure with the observed limit of rotational measure
${\rm RM}=5~{\rm rad}~{\rm m}^{-2}$ at $z=2.5$ (Kronberg 1994) due to the
intergalactic magnetic field.
We performed this calculation using the data of the simulation in Figure.
The result indicates that, with the present value of the observed limit
in rotational measure, the existence of magnetic field of $\lsim1\mu{\rm G}$
in filaments and sheets can not be ruled out (Ryu, Kang, \& Biermann 1997).
It is interesting to notice that the equipartition magnetic field strength
in filaments and sheets, $B=0.77h\sqrt{T_7}\sqrt{\rho_b/\rho_c}~\mu{\rm G}$,
is close to this limit.
Here, $T_7$ is the temperature in unit of $10^7{\rm K}$ and $\rho_b/\rho_c$
is the baryonic density in unit of critical density.

One interesting implication of such strong magnetic field
in filaments and sheets is its effects on the propagation of high energy
cosmic rays through the universe.
The discoveries of several reliable events of high energy cosmic rays at
an energy above $10^{20}{\rm eV}$ raise questions about their origin and
path in the universe, since their interaction with the cosmic microwave
background radiation limits the distances to their sources to less than
100 Mpc, perhaps within our Local Supercluster.
In Biermann, Kang \& Ryu (1996), we noted that if the magnetic field of
$\sim1\mu{\rm G}$ or less exists inside our Local Supercluster and there
exist accretion flows infalling toward the supergalactic plane, it is
possible that the high energy cosmic rays above the so-called GZK cutoff
$(E>5\times10^{19}{\rm eV})$ can be focused in the direction perpendicular
to the supergalactic plane, analogously but in the opposite direction to
solar wind modulation.
This would explain naturally the correlation between the arrival direction
of the high energy cosmic rays and the supergalactic plane.
Also, focusing means that for all the particles captured into the sheets,
the dilution with distance $d$ is $1/d$ instead of $1/d^2$, increasing the
cosmic ray flux from any source appreciably with respect to the
three-dimensional dilution.

\begin{figure}
\vspace{-0.25truein}
\epsfysize=3.9in\epsfbox[-20 250 530 795]{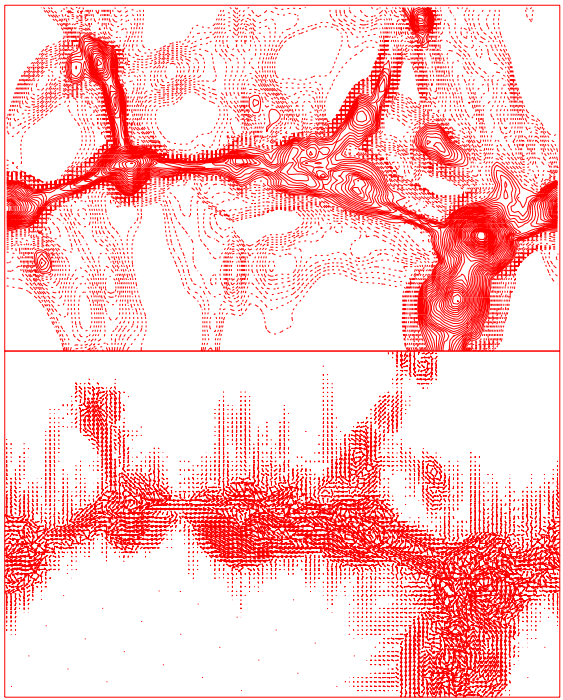}
\vspace{-0.2truein}
\caption{Two-dimensional cut of the simulated universe.
The plot shows a region of $32h^{-1}\times20h^{-1}{\rm Mpc}^2$ with
a thickness of $0.25 h^{-1}{\rm Mpc}$, although the simulation was
done in a box of $(32 h^{-1}{\rm Mpc})^3$ volume.
The upper panel shows baryonic density contours, and the lower panel
shows magnetic field vectors.~~~~~~~~~~~~~~~~~~~~~~~~~~~~~~~~~~~~~~~
~~~~~~~~~~~~~~~~~~~~~~~~~~~~~~~~~~~~~~~~~~~~~~~~~~~~~~~~~~~~~~~~~~~~}
\vspace{-0.2truein}
\end{figure}

\end{document}